\documentclass[%twocolumn,pra,aps,
onecolumn,nofootinbib,superscriptaddress]{revtex4-2}%\hoffset-2.5cm\topmargin-1.5cm\headheight0cm\headsep0cm\textheight25cm\textwidth19cm\footskip1cm
\usepackage{amsfonts}
\usepackage{amssymb}
\usepackage{bbm}
\usepackage{mathrsfs}
\usepackage{amsthm}
\usepackage{graphicx}
\usepackage{amsmath,nicematrix}
\usepackage{color}
\usepackage{tikz}
\usepackage[utf8]{inputenc}
\usepackage{empheq}

\newtheorem{Proposition}{Proposition}
\newtheorem{thm}{Theorem}
\newtheorem{con}{Conjecture}

\newtheorem{cor}{Corollary}
\newtheorem{lemma}{Lemma}
\theoremstyle{definition}

\newtheorem{fact}{Fact}[section]

\newcommand{\Tr}[0]{\mathrm{Tr}}

\newcommand{\bei}{\begin{itemize}}
\newcommand{\eei}{\end{itemize}}
\newcommand{\ket}[1]{|#1\rangle}
\newcommand{\bra}[1]{\langle#1|}

\newcommand{\ke}[1]{|#1\rangle}

\def\<{\langle}
\def\>{\rangle}
\newcommand{{\Cn}}{{\mathbb{C}^4}}
\newcommand{{\CN}}{{\mathbb{C}^{2n}}}
\newcommand{{\BC}}{{\mathcal{B}(\mathbb{C}^n)}}
\newcommand{{\BBC}}{{\mathcal{B}(\mathbb{C}^{2n})}}

\begin{document}

\title{Optimizing positive maps in the matrix algebra $M_n$}
%Results on optimization of a class of positive maps   in $\mathbb{C}^n \otimes \mathbb{C}^n$}

\author{Anindita Bera,  Gniewomir Sarbicki and Dariusz  Chru\'sci\'nski\\
\emph{Institute of Physics, Faculty of Physics, Astronomy and Informatics,  Nicolaus Copernicus University,
Grudzi\c{a}dzka 5/7, 87--100 Toru\'n, Poland}}

%\author{Anindita Bera}
%\affiliation{Institute of Physics, Faculty of Physics, Astronomy and Informatics,
%Nicolaus Copernicus University, Grudzi\c{a}dzka 5/7, 87--100 Toru{\'n}, Poland}
%\author{Filip A. Wudarski}
%\affiliation{Institute of Physics, Faculty of Physics, Astronomy and Informatics,
%Nicolaus Copernicus University, Grudzi\c{a}dzka 5/7, 87--100 Toru{\'n}, Poland}
%\affiliation{Quantum Artificial Intelligence Lab. (QuAIL), Exploration
%Technology Directorate, NASA Ames Research Center,
%Moffett Field, CA 94035, USA}
%\affiliation{USRA Research Institute for Advanced Computer Science
%(RIACS), Mountain View, CA 94043, USA}
%\author{Gniewomir Sarbicki}
%\affiliation{Institute of Physics, Faculty of Physics, Astronomy and Informatics,
%Nicolaus Copernicus University, Grudzi\c{a}dzka 5/7, 87--100 Toru{\'n}, Poland}
%\author{Dariusz Chru{\'s}ci{\'n}ski}
%\affiliation{Institute of Physics, Faculty of Physics, Astronomy and Informatics,
%Nicolaus Copernicus University, Grudzi\c{a}dzka 5/7, 87--100 Toru{\'n}, Poland}

\begin{abstract}
We present an optimization procedure for a seminal class of positive maps $\tau_{n,k}$ in the algebra of $n \times n$ complex matrices introduced and studied by Tanahasi and Tomiyama, Ando, Nakamura  and Osaka. Recently, these maps were proved to be optimal whenever the greatest common divisor $GCD(n,k)=1$. We attain a general conjecture how to optimize a map $\tau_{n,k}$ when $GCD(n,k)=2$ or 3. For $GCD(n,k)=2$, a series of analytical results are derived and for $GCD(n,k)=3$, we provide  a suitable numerical analysis.
\end{abstract}

\maketitle

\section{Introduction}

%\subsection{The family of positive maps}

%Consider $n$-dimensional complex Hilbert space

Let $M_n$ denote a matrix algebra  $n \times n$ matrices over the complex field $\mathbb{C}$. There is a natural convex cone $\mathcal{P}_n \in M_n$ consisting of positive elements in $M_n$: $X \geq 0$ if and only if $X = Y^\dagger Y$ for some $Y \in M_n$. A linear map $\Phi : M_n \to M_m$ is called positive if $\Phi(\mathcal{P}_n) \subset \mathcal{P}_m $  \cite{P1,P2,P3,P4,P5,P6,P7,P8}, i.e. the map preserves convex cones of positive elements.   
Such maps form another convex cone $\mathcal{P}_{n,m}$ in the vector space of all linear maps from $M_n$ to $M_m$. 
Interestingly, in spite of the considerable effort a cone $\mathcal{P}_{n,m}$ is still rather poorly understood (for some recent works see \cite{Q9,Q10,Q11,Q12,Q13,osaka19,kye16,osaka21,Cho1,Girard-kye,Q14,Q15,Q17,kye22}). It should be stressed that positive maps play an important role both in physics and mathematics providing generalization of $*$-homomorphisms, Jordan homomorphisms and conditional expectations. Moreover, they provide a powerful tool for characterizing quantum entanglement \cite{Q18,Q14,osaka21} and hence plays a key role in various aspects of quantum information theory \cite{Q19}.

One calls a positive map $\Phi : M_n \to M_m$ extremal if it defines an extremal element of the convex cone $\mathcal{P}_{n,m}$, that is for any positive map $\Lambda$ a map $\Phi - \Lambda$ is no longer  positive unless $\Lambda = \lambda \Phi$ with $\lambda< 1$. One calls a positive map $\Phi$ to be optimal \cite{Lew} if $\Phi - \Lambda_{\rm CP}$ is no longer a positive map, where $\Lambda_{\rm CP}$ is an arbitrary completely positive map. Clearly, any extremal map is necessarily optimal. However, the converse needs not be true. A well known example is provided by so called reduction map $R_n : M_n \to M_n$ defined by
\begin{equation}\label{Rn}
    R_n(X) =  \mathbf{I}_n {\rm Tr}\,X - X ,
\end{equation}
which is known to be optimal for any $n \geq 2$ but it is extremal only for $n=2$ \cite{Q14}. A key result concerning the structure of optimal positive maps was derived in \cite{Lew,philip22}.

\begin{thm} Let $\Phi \in \mathcal{P}_{n,m}$ and consider a set of product vectors $x_i \otimes y_i \in \mathbb{C}^n \otimes \mathbb{C}^m$ satisfying the following condition
\begin{equation} \label{SPAN}
    \< y_i ,\Phi(\overline{x}, \overline{x}_i^\dagger) y_i\> = 0 .
\end{equation}
If vectors  $\{x_i \otimes y_i\}$ span $ \mathbb{C}^n \otimes \mathbb{C}^m $, then the map $\Phi$ is optimal.    
\end{thm}
A positive map for which one can find a set of product spanning vectors satisfying (\ref{SPAN}) is said to have a spanning property. It should be stressed that the above condition is sufficient but not necessary for optimality. A well-known example of a positive optimal map without a spanning property is provided by a seminal Choi non-decomposable map in $M_3$. This problem was reviewed in great detail by Kye \cite{Q12}.   

If a positive map $\Phi$ is not optimal, one may optimize it by subtracting a certain completely positive map such that the resulting map becomes optimal after subtraction. However, in general, finding such a completely positive map is a challenging problem. In this paper, we study the optimization problem for a well-known class of positive  maps $\tau_{n,k}:M_n \to M_n$ ($k=1,2, \ldots,n-1$)  \cite{Os1,Os1a,Os2,Os3}
\begin{equation}
\tau_{n,k}(X)=(n-k) \varepsilon(X)+\sum_{i=1}^{k} \varepsilon\big(S^i X S^{\dagger i}\big)-X, 
\end{equation} 
where  $\varepsilon: M_n \to M_n$ is the canonical projection of $M_n$ to the diagonal part
 
\begin{equation}
\varepsilon(X)=\sum_{i=0}^{n-1} \Tr[X e_{ii}] e_{ii},
\end{equation}
with $e_{ij}$ being matrix units, and $S$ denotes a cyclic permutation  

\begin{equation} \label{unit_shift}
Se_i=e_{i+1},  \ \ \ \ \ (\mbox{mod}~n), 
\end{equation}
for $i=0,1,\ldots, n-1$, where $\{e_i\}$ stands for a canonical orthonormal basis in  $\mathbb{C}^n$. Actually, $\tau_{n,n-1} = R_n$ and it was proved \cite{Os2} that for $k =1,\ldots,n-2$ the map $\tau_{n,k} = R_n$ is indecomposable (even atomic meaning that it cannot be decomposed into a sum of 2-positive and 2-copositive maps).  
In particular, $\tau_{3,1}$ recovers the Choi indecomposable map in $M_3$. 

In a recent paper \cite{aniLAA} it was shown that $\tau_{n,k}$ is optimal only if the greatest  common divisor $GDC(n,k)=1$. If $d:= GDC(n,k)> 1$, then  $\tau_{n,k}$ can be optimized by subtracting a completely positive map of the following form $\lambda_{\ket{v}} H_{|v\rangle} (X)$, where
\begin{equation}
H_{|v\rangle}(X) = P_{\ket{v}} \circ X, 
\end{equation}
and $\circ$ denotes the Hadamard product. $P_{|v\rangle} := |v\rangle \langle v| $ denotes the projector onto the vector $|v\rangle$ which is a linear combination of normalised, orthogonal vectors $|v_r\rangle$:
\begin{equation}
\label{vectors}
    |v_r\rangle = \frac{1}{\sqrt{n}} \Big(1, \omega^{\frac{n}{d}r},(\omega^{\frac{n}{d}r})^2, \ldots,(\omega^{\frac{n}{d}r})^{n-1}\Big)^T, \ \ \ \mbox{with}~~ \omega=e^{2\pi i/n}, ~~\mbox{and} ~~ r=1,2,\ldots, d-1 ,
\end{equation}
that is, one considers a map
\begin{equation}
    \tau_{n,k} - \lambda_{\ket{v}} H_{|v\rangle} ,
\end{equation}
where $|v\rangle = \sum_{r=1}^{d-1} \alpha_r |v_r\rangle$, together with $\sum_r |\alpha_r|^2=1$, and the subtraction parameter $\lambda_{\ket{v}} \geq 0$ depends upon $|v\rangle$. 

In particular, if $d = 2$, one has $\ket{v} = \ket{v_1}$ and
\begin{equation}
    |v_1\rangle = \frac{1}{\sqrt{n}} \Big(1, \omega^{\frac{n}{2}},(\omega^{\frac{n}{2}})^2, \ldots,(\omega^{\frac{n}{2}})^{n-1}\Big)^T =
    \frac{1}{\sqrt{n}} \Big(1,-1,1,-1, \ldots, 1,-1 \Big)^T.
\end{equation}

Let $\lambda_{\ket{v}}^{max}$ be the largest value of $\lambda_{\ket{v}}$, such that the map $\tau_{n,k} - \lambda_{\ket{v}} H_{|v\rangle}$ is positive.

It was shown \cite{aniLAA} that for $d=2$ one has 

$$\lambda_{\ket{v}} \leq \lambda_{\ket{v}}^{max} \le n-k . $$  
Hence, it is clear that

\begin{fact}
\label{fact1}
    For $GCD(n,k)=2$, if the map $\tau_{n,k} - (n-k)H_{|v\rangle}$ is positive, then it is also optimal.
\end{fact}
In this paper we propose the following

\begin{con} \label{CON1}
    For $GCD(n,k)=2$ the map $\tau_{n,k} - (n-k)H_{|v\rangle}$ is always positive (and hence optimal).
\end{con}
In what follows we prove positivity of  the map $\tau_{n,k} - (n-k)H_{|v\rangle}$  for several pairs $(n,k)$ (cf. Corollary 1 and 2). The general proof is, however, still missing.

%and conjecture, that $n-k$ is the maximal value of $\lambda_{\ket{v_1}}$, and hence that the map 
%\begin{equation}
%\tau_{n,k}(X) - H_{|v\rangle}(X)
%\end{equation}
%is optimal. In the following paper we prove this conjecture for a subset of pairs $(n,k)$ such that $GCD(n,k)=2$. 

For $GCD(n,k)=3$, one has $\ket{v} = \alpha \ket{v_1} + \beta \ket{v_2}$ where  $\alpha, \beta \in \mathbb{C}$ s.t. $|\alpha|^2+|\beta|^2=1$, and
\begin{align}
    \ket{v_1} & =  \frac{1}{\sqrt{n}}  \Big(1, \omega^{\frac{n}{3}},(\omega^{\frac{n}{3}})^2 \ldots,(\omega^{\frac{n}{3}})^{n-1}\Big)^T,
    \label{v1_vec}
    \\
    \ket{v_2} & =  \frac{1}{\sqrt{n}}  \Big(1, \omega^{\frac{2n}{3}},(\omega^{\frac{2n}{3}})^2 \ldots,(\omega^{\frac{2n}{3}})^{n-1}\Big)^T.
    \label{v2_vec}
\end{align}
The set of vectors $\ket{v}$ constitutes a three-dimensional sphere, but while vectors differing by a phase give rise to the same projector, the set of $P_{\ket{v}}$ 
is the Bloch ball $S^3 / S^1$.
%may be parameterized by a point on a unit sphere $|\alpha|^2+|\beta|^2=1$  (Bloch sphere) since $|v\rangle$ is defined only up to a phase factor $e^{i \phi}$ with $\phi \in \mathbb{R}$. 

\begin{con}  \label{CON2}
    If $GCD(n,k)=3$, then for the optimized map
        $\tau_{n,k}^{\lambda,\alpha,\beta}=\tau_{n,k} -\lambda_{\ket{v}} H_{\ket{v}}$
where $\ket{v}=\alpha \ket{v_1}+\beta\ket{v_2}$  one has

$$\lambda_{\ket{v}}^{max} \in \big[n-k, n-k+\frac{n-3}{n-\frac{2}{3} k}\big] . $$
Moreover, 
\begin{empheq}[left=\empheqlbrace]{align}
        \label{range_lambda}
        \lambda_{\ket{v}}^{max} & =n-k ~\mbox{if}~ \alpha=0 ~\mbox{or} ~~ \beta=0, \\
        \lambda_{\ket{v}}^{max} & =  n-k+\frac{n-3}{n-\frac{2}{3} k} ~\mbox{if}~ |\alpha| =|\beta|.
    \end{empheq}
\end{con}
We provide a suitable numerical analysis showing how $\lambda_{\ket{v}}^{max}$ depends on the point of the Bloch sphere. Finally, in the conclusions, we provide a discussion about the general case $GCD(n,k)=d$.

%The corresponding entanglement witness $\mathbf{W}_{n,k}$ reads as follows
%
%\begin{equation}
 %    \mathbf{W}_{n,k} = (n-k) \sum_{i=0}^{n-1} e_{ii} \otimes e_{ii} + %\sum_{m=1}^k e_{ii} \otimes e_{i+m,i+m} - nP^+_n ,
%\end{equation}
%where $P^+_n$ denotes the projector onto the canonical maximally entangled state
%
%\begin{equation}
%     P^+_n  = \frac 1n \sum_{i,j=0}^{n-1} e_{ij} \otimes e_{ij} .
%\end{equation}
%It has a very simple block structure
%
%\begin{equation}
%     \mathbf{W}_{n,k} =  \sum_{i,j=0}^{n-1} e_{ij} \otimes W_{ij} ,
%\end{equation}
%where $W_{ij} = - e_{ij}$ for $i\neq j$, and the diagonal blocks reads $W_{ii} = S^i W_{00} S^{i\dagger}$ with
%
%\begin{equation}
%    W_{00} = {\rm Diag}[n-k-1,\underbrace{1,\ldots,1}_k,\underbrace{0,\ldots,0}_{n-k-1}] . 
%\end{equation}
%In particular when $k=n-2$ one finds
%
%\begin{equation}
%    W_{00} = {\rm Diag}[\underbrace{1,\ldots,1}_{n-1},0] , 
%\end{equation}
%which provides a generalization of the Choi witness from $n=3$. 

%Note that the matrix in the above equation is a diagonal one, %so except the diagonal terms, all the other terms are zero.

%\subsection{Optimality and optimisation of maps}
%From Ref.~\cite{aniLAA}, it has been  shown that the map $\tau_{n,k}$ is optimal for $GCD(n,k)=1$. However, if  $GCD(n,k) =d \geq 2$, then $\tau_{n,k}$ is not optimal but can be optimised 

%let us denote by $P_{v_j}$,  the corresponding projector.
%The subspace  $\mathrm{span}\{v_j\}$ is the kernel of $M$, and the possibility of corrections to the map cannot be excluded for $|\alpha\rangle$ from this subspace. 

\section{Optimization for $GCD(n,k)=2$}

If $GCD(n,k)=2$ then  $n,k$ are necessarily even. In this case positivity of  $\tau_{n,k}(X)-  (n-k) H_{|v_1\rangle}$ may be reformulated as follows

\begin{thm}
\label{main_th}
%Let $2|n,k$, $\lambda=n-k$ and $\circ$ denotes the Hadamard product. 
The map
%If  $2|n,k$, then the map:
% in Eq.~(\ref{map1}) can always be optimized by a rank one projector $P_{v_1}$ in the following way:
\begin{equation}
\label{p1}
   % \tau_{n,k}^{\lambda} (X)=
   \tau_{n,k}(X)-  (n-k) P_{|v_1\rangle} \circ X, 
\end{equation}
%and  $P_{v_1}=|v_1 \rangle \langle v_1|$, where $|v_1 \rangle$ corresponds to the Eq.~(\ref{vectors}) for $i=1$ i.e.
%\begin{eqnarray}
%    |v_1\rangle =& \frac{1}{\sqrt{n}} \Big(1, w^{\frac{n}{d}},(w^{\frac{n}{d}})^2, \ldots,(w^{\frac{n}{d}})^{n-1}\Big) \nonumber\\
 %   =&\frac{1}{\sqrt{n}} \Big(1,-1,1,-1, \ldots, 1,-1 \Big).
%\end{eqnarray}
%and $\circ$ denotes the Hadamard product, hence
%\begin{equation}
%\lambda~P_{v_1} \circ X=\frac{\lambda}{n}
%\begin{bmatrix}
%  X_{00} & -X_{01} & \dots & (-1)^{n-2} X_{0,n-2} &  (-1)^{n-1} X_{0,n-1}  \\
%    -X_{10} & X_{11} & \dots & (-1)^{n-1} X_{1,n-2} &  (-1)^{n} X_{1,n-1}  \\
% \vdots & \vdots & \ddots & \vdots & \vdots   \\
% (-1)^{n+1}  X_{n-1,0} & (-1)^{n+2}  X_{n-2,0} & \dots &  (-1)^{n+2} X_{n-1,n-2} & % (-1)^{n+1} X_{n-1,n-1} \\
%\end{bmatrix}
%\end{equation}
is positive if and only if 
\begin{align}
 \label{p7M}
 \frac {nk}4  \min \vec u \min \vec v (
 \vec u^T A_{n,k} \vec u + \vec u^T B_{n,k} \vec v
  + 
  \vec v^T A_{n,k} \vec v + \vec v^T C_{n,k} \vec u
  ) 
%  \nonumber \\
  + 
  (\vec u^T A_{n,k} \vec u + \vec u^T B_{n,k} \vec v)
  (\vec v A_{n,k} \vec v + \vec v^T C_{n,k} \vec u)
  ~\geq~ 0 ,
 \end{align}
for all vectors $\vec u, \vec v \in \mathbbm{1}_{n/2}^{\perp} \subset \mathbb{R}^{n/2}$, that is,

$$  \sum_{i=1}^{n/2} u_i = \sum_{i=1}^{n/2} v_i = 0 .$$
Moreover, $ \min \vec u := \min_i u_i$ and $\min \vec v := \min_i v_i$. 
Three $\frac{n}{2} \times \frac{n}{2}$  matrices $A_{n,k}, B_{n,k}$ and $C_{n,k}$ are defined as follows
\begin{equation}
    A_{n,k} = (n-k) \mathbf{I}_{n/2}+\sum_{i=1}^{k/2} P_{n/2}^i,~ B_{n,k}=\sum_{i=0}^{k/2-1} P_{n/2}^i,~ C_{n,k}=\sum_{i=1}^{k/2} P_{n/2}^i,
\end{equation}
where  $P_n=\sum_{i=1}^{n} |e_i \rangle \langle e_{i+1}|$ is the cyclic permutation matrix and $P_n^i$ is the $i$-th power of  $P_n$.

\end{thm}
For the proof, cf. the Appendix \ref{appA}. As an  illustration, we consider a few simple examples of pairs $(n,k)$:
\begin{itemize}
    \item  For  $n=4$ and $k=2$ :
\begin{equation*}
    A_{4,2}=\left(
\begin{array}{cc}
 2 & 1 \\
 1 & 2 \\
\end{array}
\right)=2 \mathbf{I}_2+P_2^1,~ 
B_{4,2}=\left(
\begin{array}{cc}
 1 & 0 \\
 0 & 1 \\
\end{array}
\right)=P_2^0=\mathbf{I}_2,~
 C_{4,2}=\left(
\begin{array}{cc}
 0 & 1 \\
 1 & 0 \\
\end{array}
\right)=P_2^1.
\end{equation*}

  \item  For  $n=8$ and $k=4$ :
\begin{equation*}
   A_{8,4}=\left(
\begin{array}{cccc}
 4 & 1 & 1 & 0 \\
 0 & 4 & 1 & 1 \\
 1 & 0 & 4 & 1 \\
 1 & 1 & 0 & 4 \\
\end{array}
\right)=4 \mathbf{I}_4+P_4^1+P_4^2,~~
B_{8,4}=\left(
\begin{array}{cccc}
 1 & 1 & 0 & 0 \\
 0 & 1 & 1 & 0 \\
 0 & 0 & 1 & 1 \\
 1 & 0 & 0 & 1 \\
\end{array}
\right)=P_4^0+P_4^1,~
C_{8,4}=\left(
\begin{array}{cccc}
 0 & 1 & 1 & 0 \\
 0 & 0 & 1 & 1 \\
 1 & 0 & 0 & 1 \\
 1 & 1 & 0 & 0 \\
\end{array}
\right)=P_4^1+P_4^2.
\end{equation*}

 \item  For  $n=8$ and $k=6$ :
\begin{align*}
   A_{8,6}=\left(
\begin{array}{cccc}
 2 & 1 & 1 & 1 \\
 1 & 2 & 1 & 1 \\
 1 & 1 & 2 & 1 \\
 1 & 1 & 1 & 2 \\
\end{array}
\right)=2 \mathbf{I}_4+P_4^1+P_4^2+P_4^3,~~
&
B_{8,6}=\left(
\begin{array}{cccc}
 1 & 1 & 1 & 0 \\
 0 & 1 & 1 & 1 \\
 1 & 0 & 1 & 1 \\
 1 & 1 & 0 & 1 \\
\end{array}
\right)=P_4^0+P_4^1+P_4^2,~
\\
&
C_{8,6}=\left(
\begin{array}{cccc}
 0 & 1 & 1 & 1 \\
 1 & 0 & 1 & 1 \\
 1 & 1 & 0 & 1 \\
 1 & 1 & 1 & 0 \\
\end{array}
\right)=P_4^1+P_4^2+P_4^3.
\end{align*}
\end{itemize}

%{\bf ANINDITA: these matrices are too trivial. Pls consider higher dimensions, e.g. n=8 and k=4 and 6.}

The matrices $B_{n,k}$ and $C_{n,k}$ enjoy the following properties: 
 
 %To prove the following condition, we need the following lemma:
 \begin{lemma} \label{Darek's} $\min_{\vec x, \vec y \in \mathbbm{1}_{n/2}^{\perp}} \vec x^T B_{n,k} \vec y = \min_{\vec x,\vec y \in \mathbbm{1}_{n/2}^{\perp}} \vec x^T C_{n,k} \vec y = -\frac{nk}4 \min \vec x \min \vec y$.
 \proof Without loss of generality, let us assume that

$$  \min \vec x = \min \vec y = - 1 .$$
The aim is to show that 

\begin{equation}   \label{L!}
    \min_{\vec x, \vec y \in \mathbbm{1}_{n/2}^{\perp}} \vec x^T B_{n,k} \vec y = \min_{\vec x, \vec y \in \mathbbm{1}_{n/2}^{\perp}} \vec x^T C_{n,k} \vec  y = -\frac{nk}4 .
\end{equation}
We provide the proof for the matrix $B_{n,k}$. The proof for $C_{n,k}$ uses essentially the same arguments. One has

\begin{equation}\label{SUM}
    \vec x^T B_{n,k} \vec y = \vec x^T \vec y  + \vec x^T  P_{n/2} \vec y + \vec x^T P_{n/2}^2 \vec y + \ldots + \vec x^T P_{n/2}^{\frac k2-1} \vec y .
\end{equation}
Each element in the above sum has a form $\vec a^T \vec b$ with $ \vec{a}, \vec{b} \in \mathbbm{1}_{n/2}^{\perp}$, and $\min \vec a = \min \vec b =-1$.  
Note that
We will prove first that $\vec a^T \vec b \geq  - \frac n2 $. Let us first consider the case, when the minimum in both vectors is attained in different positions. Then one can permute the basis to write the vectors in the form:

\begin{equation}
  \vec  a = \Big(-1, 1 -(a_1+\ldots+a_{\frac n2 -2}),a_1,\ldots,a_{\frac n2-2}\Big) , \ \ \ \vec b = \Big(1 -(b_1+\ldots+b_{\frac n2-2}),-1,b_1,\ldots,b_{\frac n2-2}\Big) ,
\end{equation}
with $a_i,b_j \geq -1$. One finds

\begin{equation}
   \vec a^T \vec b = -2 + (a_1+\ldots+a_{\frac n2-2}) + (b_1+\ldots+b_{\frac n2-2}) + \sum_{i=1}^{\frac n2-2} a_i b_i .
\end{equation}
One has

\begin{equation}
    0 \leq (a_1+1)(b_1+1) + \ldots + (a_{\frac n2-2}+1)(b_{\frac n2-2}+1) = (a_1+\ldots+a_{\frac n2-2}) + (b_1+\ldots+b_{\frac n2-2}) + \sum_{i=1}^{\frac n2-2} a_i b_i + \frac n2 -2 , 
\end{equation}
and hence

\begin{equation} \label{aTb}
   \vec a^T \vec b \geq - \frac n2 . 
\end{equation}

It remains to consider the situation, when the minima of both vectors are attained in the same coordinate. Permute the basis to make it the first coordinate. Then the remaining coordinates constitute $(n/2-1)$ - dimensional vectors. Let us denote them by $\vec a_{rest}$ and $\vec b_{rest}$. Let their minima be $a_{min}$ and $b_{min}$ respectively. Obviously,  $a_{min}, b_{min} > -1$. We have:
\begin{equation}
  \vec  a = \left( -1, \frac 1{\frac n2 - 1} \vec{\mathbbm{1}}_{\frac n2 -1} - \big( a_{min} - \frac 1{\frac n2 - 1} \big) \widetilde{a}_{rest} \right)
\end{equation}
where $\widetilde{a}_{rest}$ is the variable component of $\vec a_{rest}$ normalised to: $\min \widetilde{a}_{rest} = -1$, hence satisfying  (\ref{aTb}) in the dimension lower by $1$. The vector $\vec b_{rest}$ will be represented similarly. Now we have:
\begin{align}
  \vec  a^T \vec b 
    & = 1 + \frac 1{\frac n2 - 1} + \left( a_{min} - \frac 1{\frac n2 - 1} \right) \left( b_{min} - \frac 1{\frac n2 - 1} \right) \widetilde{a}_{rest}^T \widetilde{b}_{rest}
    \nonumber \\
    & \ge 1 + \frac 1{\frac n2 - 1} + \left( -1 - \frac 1{\frac n2 - 1} \right) \left( -1 - \frac 1{\frac n2 - 1} \right) \left( -\frac n2 + 1 \right) = - \frac n2
\end{align}
and (\ref{aTb}) is valid also in all the cases.

Finally, let us observe that for the following pair of vectors

\begin{equation}
   \vec a = (-1,\frac n2-1,-1,\ldots,-1) , \ \ \ \vec b = (-1,-1,\ldots,-1,\frac n2-1) ,
\end{equation}
each separate term in (\ref{SUM}) is exactly equal to $-\frac n2$ and hence  (\ref{L!}) follows. $\square$
 
 \end{lemma}
 
 The spectrum of the cyclic permutation matrix $P_{n/2}$ reads

\begin{equation}
    {\rm spec}(P_{n/2}) = \{1\} \cup \left\{ \exp\left( \frac{2\pi}{n/2} j \right) \right\}_{j=1}^{n/2-1}.
\end{equation}
One has therefore

\begin{equation}
    {\rm spec}(C_{n,k}) = \{ k/2\} \cup \left\{ \exp\left( \frac{2\pi}{n/2} j \right) \left( 1 - \exp\left( \frac{2\pi}{n/2} \frac k2 j \right) \right)  \left( 1 - \exp\left( \frac{2\pi}{n/2} j \right)^{-1} \right) \right\}_{j=1}^{n/2-1}.
\end{equation}
The eigenvalues $c_{n,k}^{(j)}$ of $(C_{n,k}+C_{n,k}^T)/2$ are the real parts of the eigenvalues of $C_{n,k}$ and hence
\begin{equation}
\label{cnkj}
   c_{n,k}^{(j)} =
    \begin{cases}
      \frac{1}{2} \left( \frac{\sin\left( \frac{2\pi}n (k+1) j \right)}{\sin\left( \frac{2\pi}n j \right)} - 1 \right), & \text{$j \in \{ 1,\ldots, \frac n2-1\}$   },\\
     \frac k2 & \text{ $j=0$ }.
    \end{cases}       
\end{equation}

%\begin{equation} 
 %   c_{n,k}^{(j)} =\frac{1}{2} \left( \frac{\sin\left( \frac{2\pi}n (k+1) j \right)}{\sin\left( \frac{2\pi}n j \right)} - 1 \right).
%\end{equation}
Let $c_{n,k}^{min}$ and $c_{n,k}^{max}$ be the smallest and the largest eigenvalue of $(C_{n,k}+C_{n,k}^T)/2$. One has the following

 \begin{fact} 
 \label{fact1}
 For $k \ge 4$ we have the following estimations: 
 $c_{n,k}^{max} \leq \frac{k}{2}$
 and
 \begin{displaymath}
    c_{n,k}^{min} \geq ( -\big[\sin(\frac{\pi}{k+1})\big]^{-1} - 1 ) / 2 \geq 
  - \frac 12 (\mu(k+1) + 1), 
 \end{displaymath}
where

\begin{equation}
    \mu = \left(\frac 52  \sqrt{\frac 12 (5-\sqrt{5})} \right)^{-1} \approx 0.34026.
\end{equation}

 \end{fact}
\emph{Proof:} For $c_{n,k}^{max}$,  we have
\begin{equation}
    \frac{\sin\left( \frac{2\pi}n (k+1) j \right)}{\sin\left( \frac{2\pi}n j \right)} \leq k+1.
\end{equation}
This implies $ c_{n,k}^{max} \leq \frac{1}{2} (k+1-1) \leq \frac{k}{2}.$

For $k \geq 4$ from Eq.~\eqref{cnkj}, the argument of the $sine$ function can go from $0$ to $\frac{\pi}{5}$. Note that $\sin\frac{\pi}{5}=\sqrt{\frac{5}{8}-\frac{\sqrt{5}}{8}}$. 
Since $sine$ function is concave in $[0,\pi]$, then from Jensen's inequality
\begin{equation}
    \forall x \in[0,\pi/5],~~ \sin x \geq \frac{\sin \frac{\pi}{5}}{\frac{\pi}{5}} x.
\end{equation}
Therefore, $\sin\frac{\pi}{k+1} \geq \frac{\sqrt{\frac{5}{8}-\frac{\sqrt{5}}{8}}}{\frac{\pi}{5}} \frac{\pi}{k+1}=\frac{5}{k+1} \sqrt{\frac{5}{8}-\frac{\sqrt{5}}{8}}$. Now, for $ c_{n,k}^{min}$, we have

\begin{eqnarray} 
 \min_{X \in [0,\pi]} \frac{\sin\left((k+1) X \right)}{\sin X} &=&  \min_{\substack{X \in [0,\pi] \\ \sin((k+1)X) <0}} \frac{\sin\left((k+1) X \right)}{\sin X} 
 \geq  \min_{\substack{X \in [0,\pi] \\ \sin((k+1)X) <0}} \sin\left((k+1) X \right) \times \max_{\substack{X \in [0,\pi] \\ \sin((k+1)X) <0}} \frac{1}{\sin X} \nonumber\\
&=& (-1) \times \frac{1}{\sin\frac{\pi}{k+1}}
  \geq -\frac{k+1}{5 \sqrt{\frac{5}{8}-\frac{\sqrt{5}}{8}}} =\mu (k+1).    %\nonumber\\
\end{eqnarray}
Hence, $ c_{n,k}^{min} \geq -\frac{1}{2} \Big( \big[\sin\frac{\pi}{k+1}\big]^{-1} + 1 \Big) = -\frac 12 \big(\mu (k+1) + 1\big)$.
 
 \begin{Proposition}\label{lemma_cj}

    %If $\forall j \in \{1, \dots, \frac{n}{2}-1 \},~ (n-k + c_{n,k}^{min})(n-k + c_{n,k}^{(j)}) \ge \frac {(n-2)k}4 |c_{n,k}^{(j)}|$, 
    If $k=2$ or
    \begin{empheq}[left=\empheqlbrace]{align}
        \label{c_max_cond}
        (n-k+c_{n,k}^{min})(n-k+c_{n,k}^{max}) & \ge \frac k4 (n-2) c_{n,k}^{max}, \\
        (n-k+c_{n,k}^{min})^2 & \ge - \frac k4 (n-2) c_{n,k}^{min}.
        \label{c_min_cond}
    \end{empheq}
    then the condition (\ref{p7M}) holds (and hence the map $ \tau_{n,k}(X)-  (n-k) H_{|v_1\rangle}$
    is positive).
     \end{Proposition} 
For the proof cf. the Appendix \ref{appB}. Consider now specific pairs $(n,k)$ for which assumptions of Proposition~\ref{lemma_cj} are satisfied. We present the following admissible classes:

%The spectrum of the cyclic permutation matrix $P$ are $\left\{ \exp\left( \frac{2\pi}{n/2} j \right) \right\}_{j=0}^{n/2-1}$. The spectrum of $C = \sum_{i=1}^{k/2} P^i$ is then $\left\{ \exp\left( \frac{2\pi}{n/2} j \right) \left( 1 - \exp\left( \frac{2\pi}{n/2} \frac k2 j \right) \right) / \left( 1 - \exp\left( \frac{2\pi}{n/2} j \right) \right) \right\}_{j=0}^{n/2-1}$. The eigenvalues $c_j$ of $(C+C^T)/2$ are the real parts of the eigenvalues of $C$, hence:
%\begin{equation}
%    c_j = \left( \frac{\sin\left( \frac{2\pi}n (k+1) j \right)}{\sin\left( \frac{2\pi}n j \right)} - 1 \right) / 2.
%\end{equation}

\begin{cor}   \label{COR1}
\label{ani}
If $n \in \{ 2k, 2k+2, 2k+4 \}$ then the assumptions of Proposition~\ref{lemma_cj} are satisfied (and hence the map $\tau_{n,k} - (n-k)H_{|v\rangle}$ is positive).
\end{cor}

\begin{figure}[h!]
    \centering
    \includegraphics[width=\textwidth]{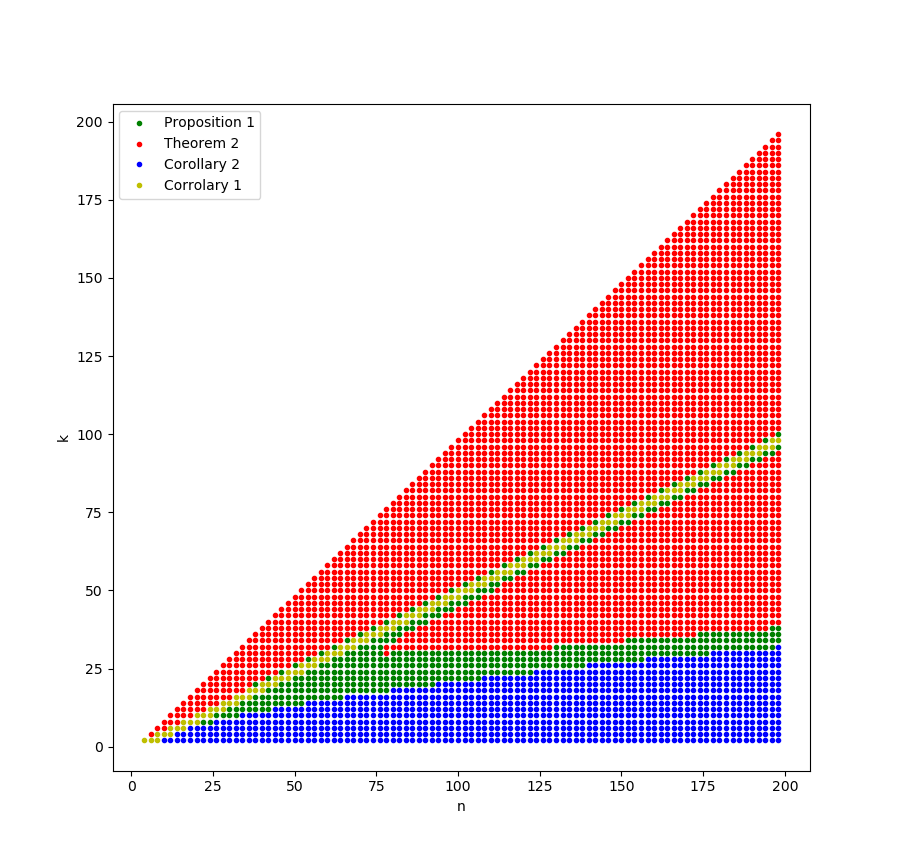}
    \caption{Pairs $(n,k)$, $n,k$-even satisfying: Corollary~\ref{ani} (yellow points), Corollary~\ref{gni} (blue points), Proposition~\ref{lemma_cj} (green points) and Theorem~\ref{main_th} (red points).}
    \label{fig:cor}
\end{figure}

\begin{cor}   \label{COR2}
  \label{gni}
  If $n,k \ge 4$ and $16 n \ge 2k^2+8(\mu+4)k+4(\mu+1)-27$, then the assumptions of Proposition~\ref{lemma_cj} are satisfied (and hence the map $\tau_{n,k} - (n-k)H_{|v\rangle}$ is positive).
\end{cor}
See Appendices \ref{AppC} and \ref{AppD} for the proofs. The Figure~\ref{fig:cor} presents the pairs $(n,k)$, $n,k$-even satisfying: Corollary~\ref{ani} (yellow points), Corollary~\ref{gni} (blue points), Proposition~\ref{lemma_cj} (green points, numerical results) and Theorem~\ref{main_th} (red points, numerical results).

\begin{figure}
\centering
\includegraphics[width=\textwidth]{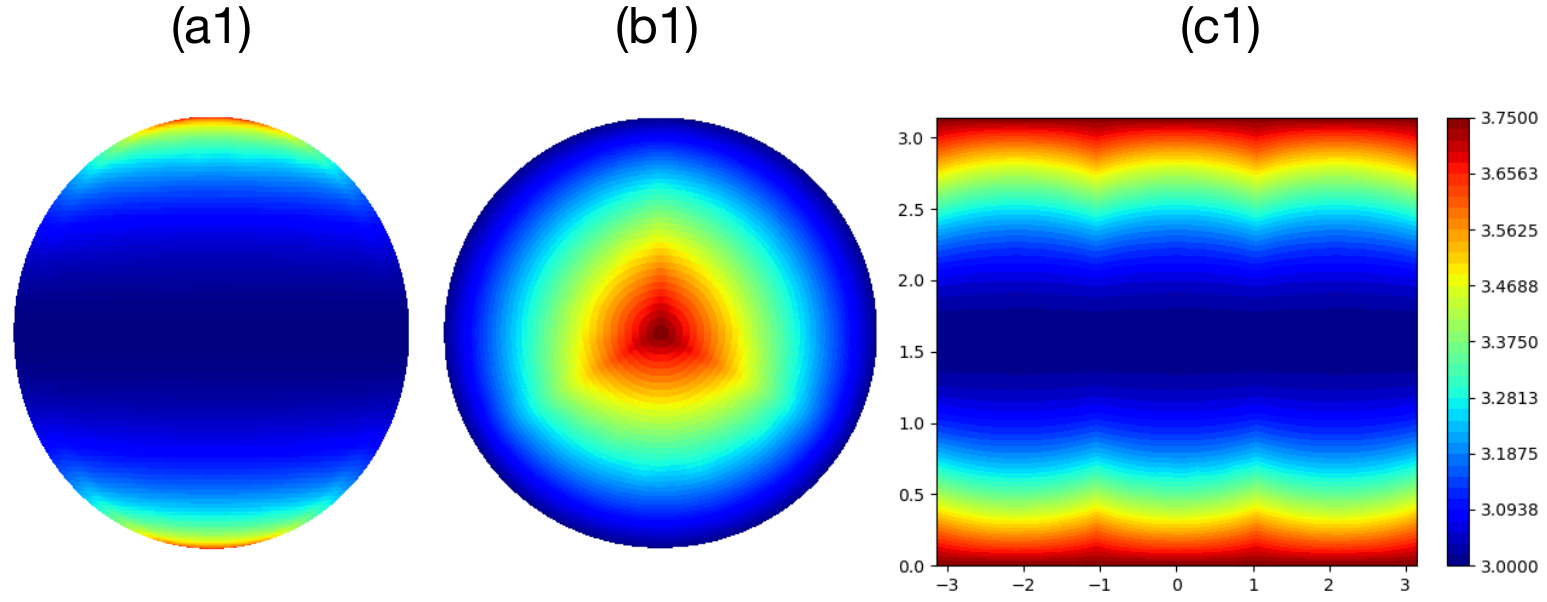}
\includegraphics[width=\textwidth]{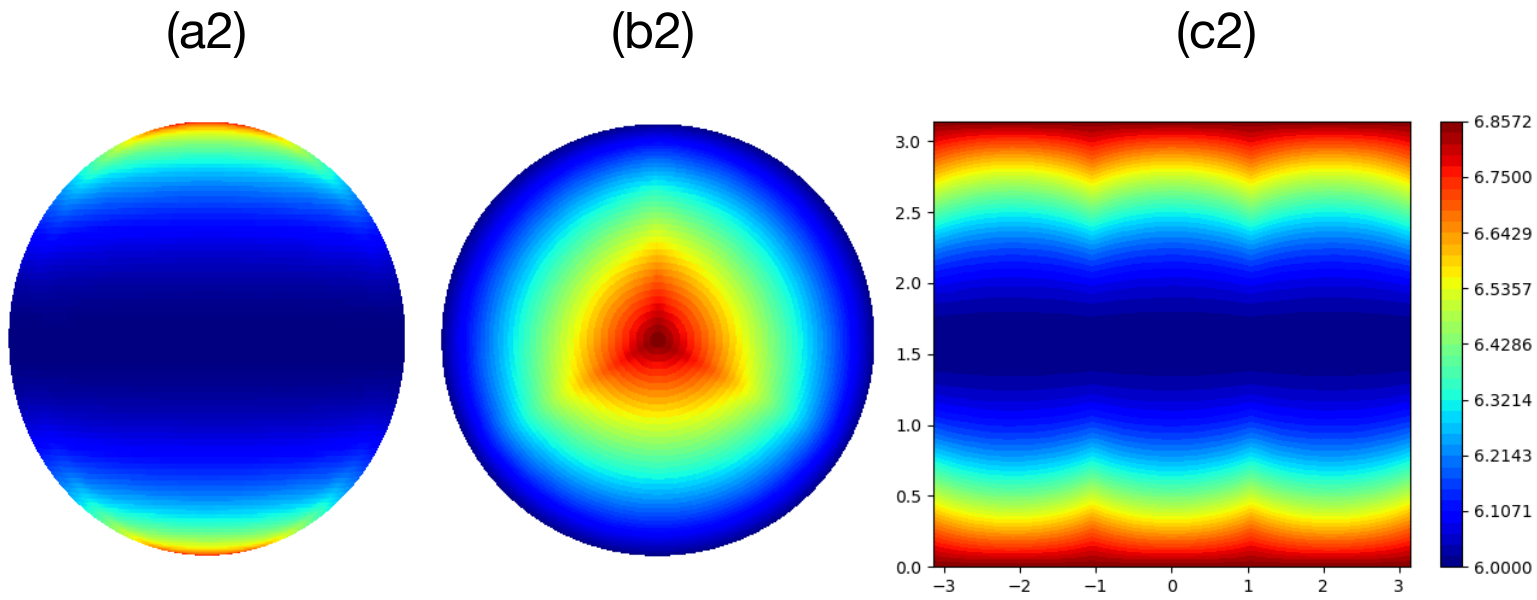}
\includegraphics[width=\textwidth]{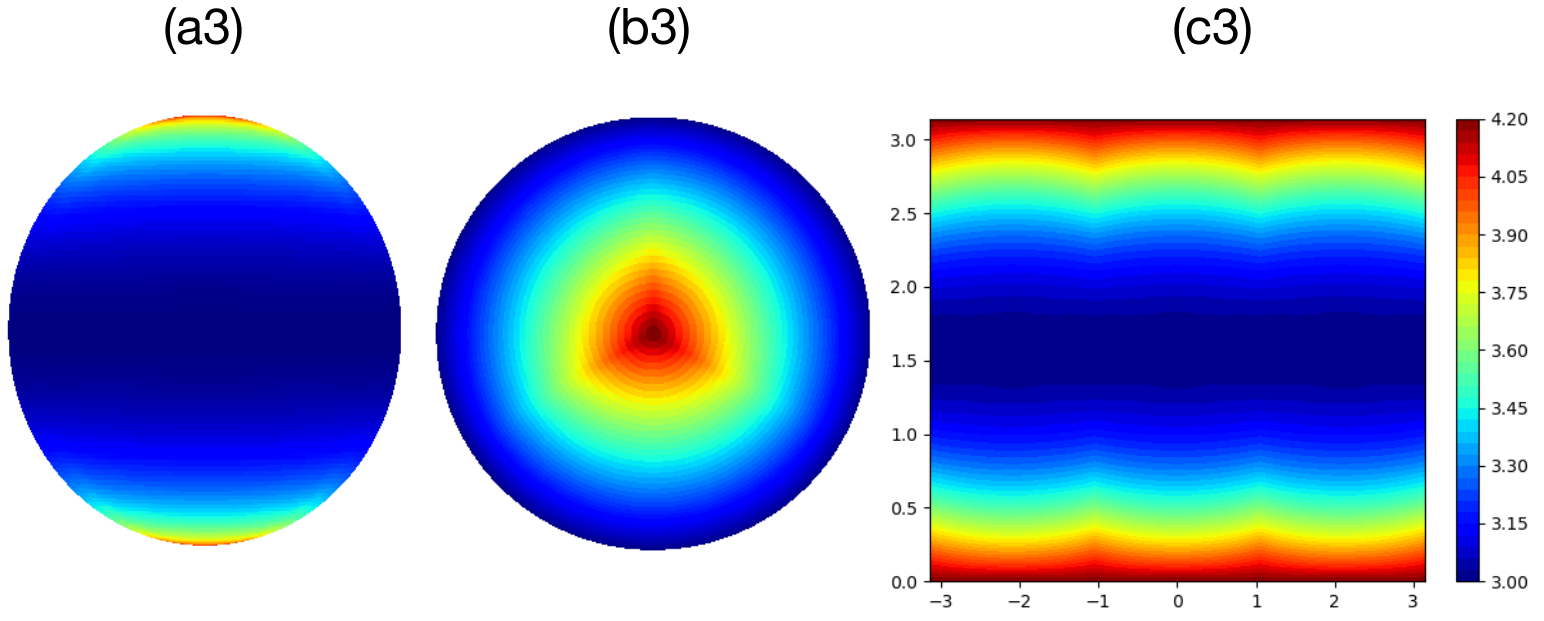}
\caption{(ai) and (bi) represent the side and top views on the Bloch spheres of 1D projectors supported on the 2D subspaces $\mathrm{span}\{v_1,v_2\}$ where colour denotes the value of $\lambda_{\ket{v}}^{max}$.
(ci) corresponds to the contour plots of $\lambda_{\ket{v}}^{max}$ in  the spherical co-ordinates  $\phi$ ($x$ axis) and  $\theta$ ($y$-axis), where $\phi \in [-\pi,\pi]$ and $\theta \in [0,\pi]$. 
%Here the colors exhibit the maximal amount of the projector $(\lambda P_{\alpha v_1+\beta v_2})$ that can be subtracted without destroying the positivity, to optimize the map.
  %Here i=1, 2, 3. 
  %The colorbar expresses the ranges of different $\lambda$ values.
First, second and third rows correspond to $(n,k) = (6,3), (9,3)$ and $(9,6)$ respectively.
%, which shows that $GCD(n,k)=3$. 
}
\label{fig1}
\end{figure}
In a range of pairs $(n,k)$, we have proven that  maps $\tau_{n,k} -\lambda_{\ket{v}} H_{\ket{v}}$ are positive for $\lambda_{\ket{v}} = n-k$. Hence, due to the Fact~\ref{fact1}, if $GCD(n,k)=2$, these maps are also optimal.

\section{Optimization for $GCD(n,k)=3$}

In this section we  formulate the following conjecture based on our numerical observation. The details of calculations are in the Appendix \ref{AppE}.

%The numerical analysis we have performed, lets us to formulate the following:
\begin{con}
    If $GCD(n,k)=3$, then for the optimized map
        $\tau_{n,k}^{\lambda,\alpha,\beta}=\tau_{n,k} -\lambda_{\ket{v}} H_{\ket{v}}$
where $\ket{v}=\alpha \ket{v_1}+\beta\ket{v_2}$ ($\ket{v_1}$ and $\ke{v_2}$ are given in \eqref{v1_vec} and \eqref{v2_vec}) one has

$$\lambda_{\ket{v}}^{max} \in \big[n-k, n-k+\frac{n-3}{n-\frac{2}{3} k}\big] , $$
 where
\begin{empheq}[left=\empheqlbrace]{align}
        \label{range_lambda}
        \lambda_{\ket{v}}^{max} & =n-k ~\mbox{if}~ \alpha=0 ~\mbox{or} ~~ \beta=0, \\
        \lambda_{\ket{v}}^{max} & =  n-k+\frac{n-3}{n-\frac{2}{3} k} ~\mbox{if}~ |\alpha| =|\beta|.
    \end{empheq}
\end{con}

In Figure~\ref{fig1}, we plot the value of $\lambda_{\ket{v}}^{max}$ on the Bloch sphere for $(n,k) = (6,3), (9,3)$ and $(9,6)$. We observe the perfect agreement of the range of $\lambda_{\ket{v}}^{max}$ with the conjecture. Also the maximal and minimal values of $\lambda_{\ket{v}}^{max}$ are attained in the predicted points (poles and equator respectively).

We observe that the graphs are symmetric w.r.t. rotation: $\phi \mapsto \phi + 2\pi/3$. Actually we are able to prove the below proposition:

{\bf Proposition 1:}
\emph{For  $GCD(n,k)=d=3$, 
% following the Hypothesis 2, 
%the map  $\tau_{n,k}$  in Eq.~(\ref{map1}) can always be optimized by a projector of linear combination of two vectors i.e. $\alpha |v_1 \rangle+\beta |v_2 \rangle$. % 
a rotation on the parameter $\beta$ with the phase $e^{2 \pi i/3}$ in the Bloch sphere  provides a  map related to the original one given in Eq.~(\ref{map1}) by 
composition with unitary channels on input and output.}

\emph{Proof}: Let us write the linear combination of two vectors explicitly below in the following way:
\begin{eqnarray}
\alpha |v_1 \rangle+\beta |v_2 \rangle &=& \alpha \frac{1}{\sqrt{n}}  \Big(1, \omega^{\frac{n}{3}},(\omega^{\frac{n}{3}})^2 \ldots,(\omega^{\frac{n}{3}})^{n-1}\Big)+\beta \frac{1}{\sqrt{n}}  \Big(1, \omega^{\frac{2n}{3}},(\omega^{\frac{2n}{3}})^2 \ldots,(\omega^{\frac{2n}{3}})^{n-1}\Big) \nonumber\\
&=& \frac{1}{\sqrt{n}}\Big(\alpha+\beta, e^{\frac{2 \pi}{3} i} \alpha+e^{-\frac{2 \pi}{3}i} \beta, e^{-\frac{2 \pi}{3}i} \alpha+e^{\frac{2 \pi}{3}i} \beta, \alpha+\beta,\ldots, e^{\frac{2 \pi}{3}  (n-1) i} \alpha+e^{-\frac{2 \pi}{3} (n-1) i} \beta  \Big).
\end{eqnarray}
Now, a rotation on $\beta$ with angle $e^{\frac{2 \pi}{3}i}$ will change the linear combination mentioned above in the below form
\begin{eqnarray}
\label{modified_comb}
\alpha |v_1 \rangle+e^{\frac{2 \pi }{3}i}\beta |v_2 \rangle=\frac{1}{\sqrt{n}}\Big(\alpha+e^{\frac{2 \pi }{3}i}\beta, e^{\frac{2 \pi}{3} i} \alpha+ \beta, e^{-\frac{2 \pi}{3}i} \alpha+e^{-\frac{2 \pi}{3}i} \beta, \alpha+e^{\frac{2 \pi }{3}i}\beta,\ldots, e^{\frac{2 \pi}{3}  (n-1) i} \alpha+e^{-\frac{2 \pi}{3} (n-2) i} \beta  \Big), \nonumber\\
\mapsto \frac{1}{\sqrt{n}}\Big(e^{\frac{2 \pi }{3}i}\alpha+e^{-\frac{2 \pi }{3}i}\beta, e^{-\frac{2 \pi}{3} i} \alpha+e^{\frac{2 \pi }{3}i} \beta, \alpha+\beta, e^{\frac{2 \pi }{3}i}\alpha+e^{-\frac{2 \pi }{3}i}\beta,\ldots, e^{\frac{2 \pi}{3}  (n-2) i} \alpha+e^{-\frac{2 \pi}{3} (n-3) i} \beta  \Big). \nonumber\\
\end{eqnarray}
We obtain the second line of the above equation by multiplying a global phase of  $e^{\frac{2 \pi }{3}i}$.
Therefore clearly we can see that one can get the elements of the vector $\alpha |v_1 \rangle+e^{\frac{2 \pi }{3}i}\beta |v_2 \rangle$ by just the permutation of the elements of the vector in $\alpha |v_1 \rangle+\beta |v_2 \rangle$, i.e.
\begin{equation}
    \alpha |v_1 \rangle+e^{\frac{2 \pi }{3}i}\beta |v_2 \rangle= e^{-\frac{2 \pi }{3}i}
    Q \Big(\alpha |v_1 \rangle+\beta |v_2 \rangle\Big)=e^{-\frac{2 \pi }{3}i}
    Q |v\rangle,
\end{equation}
where  $Q$ is the block diagonal matrix with the diagonal block $\begin{pmatrix} \begin{matrix}
  0 & 1 & 0 \\
  0 & 0 & 1 \\
  1 & 0 & 0
  \end{matrix} \end{pmatrix} $.
  
%\begin{equation}
%  Q=  \begin{pmatrix}
%  \begin{matrix}
%  0 & 1 & 0 \\
%  0 & 0 & 1 \\
%  1 & 0 & 0
%  \end{matrix}
%  & \rvline & \bigzero  \\
%\hline
%  \bigzero & \rvline &
 %  \begin{matrix}
 % 0 & 1 & 0 \\
 % 0 & 0 & 1 \\
 % 1 & 0 & 0
%  \end{matrix}
%\end{pmatrix}.
%\end{equation}
Now, let us subtract the projector of $\alpha |v_1 \rangle+e^{\frac{2 \pi }{3}i}\beta |v_2 \rangle$ from the map in (\ref{map1}) and we get
\begin{eqnarray}
\tau_{n,k}^{\lambda,\alpha,\tilde{\beta}}(X) &=& D[X] -X-\lambda Q |v\rangle \langle v| Q^T \circ X \nonumber\\
&=& D[X] -X-\lambda Q (|v\rangle \langle v|\circ Q^T X Q)Q^T \nonumber\\
&=& Q D[Q^TXQ] Q^T- Q Q^T X Q Q^T- \lambda Q (|v\rangle \langle v|\circ Q^T X Q)Q^T
  %  \Tilde{\tau}_{n,k}(X)=\mbox{Diag}\Big[(n-k) X_{00}+X_{11}+\ldots+X_{kk},\ldots, (n-k) X_{n-1,n-1}+X_{00}+\ldots+X_{k-1,k-1} \Big]    -X-\lambda Q |v'\rangle \langle v'| Q^T \circ X
 %   =\mbox{Diag}\Big[(n-k) X_{00}+X_{11}+\ldots+X_{kk},\ldots, (n-k) X_{n-1,n-1}+X_{00}+\ldots+X_{k-1,k-1} \Big]  -X-\lambda Q (|v'\rangle \langle v'|\circ Q^T X Q)Q^T
 \nonumber \\
 &=& Q \tau_{n,k}^{\lambda,\alpha,\beta}( Q^T X Q ) Q^T.
\end{eqnarray}
Hence $\tau_{n,k}^{\lambda,\alpha,\tilde{\beta}} = \mathrm{Ad}_Q * \tau_{n,k}^{\lambda,\alpha,\beta} * \mathrm{Ad}_{Q^T}$ (here $\mathrm{Ad}_Q$ denotes a map $X \mapsto QXQ^T$ and $*$ denotes the composition of maps). $\square$

\section{Conclusions}

In this paper we analyzed optimization procedure for a class of seminal maps $\tau_{n,k}$. These  maps turned out to be optimal only when $GCD(n,k)=1$ \cite{aniLAA}. However, for $GDC(n,k)=d>2$, a suitable optimization procedure is required which in this case consists in subtracting a completely positive map being a Hadamard product with a suitable 1-rank projector $P_{|v\rangle}$. We performed analysis for $d=2$ and $d=3$. When $d=2$ we proposed Conjecture \ref{CON1} which is supported by a several pairs $(n,k)$, where both $n$ and $k$ are even. For $d=3$ we proposed Conjecture \ref{CON2} which is supported by a suitable numerical analysis.

Our analysis strongly suggests the following general result 

%We observe, that for $GCD(n,k)$ in any direction one can subtract $(n-k) H_{\ket{v}}$ from $\ket{\tau}_{n,k}$ without destroying its optimality, like in the case $n,k$ - even. We conjecture that it is a general fact and formulate the following:

\begin{con}
  If $GCD(n,k)=d \geq 2$,  the positive map  $\tau_{n,k}$ is not optimal but can be optimized by subtracting a completely positive map being the Hadamard product with a rank-1 projector $P_{|v\rangle}$, where

$$  |v\rangle = \sum_{r=1}^{d-1} \alpha_r |v_r \rangle , $$
together with a normalization condition $\sum_r |\alpha_r|^2=1$. The value of a subtraction parameter $\lambda_{|v\rangle} \leq n-k$ and in general it depends upon the complex parameters $(\alpha_1,\ldots,\alpha_{d-1}) \in S^{2(d-1)-1}$, i.e. $|\alpha_1|^2 + \ldots + |\alpha_{d-1}|^2=1$. Due to the fact that $|v\rangle$ is defined only up to the global phase factor a set of projectors $P_{|v\rangle}$ may parameterized by $S^{2(d-2)} = S^{2(d-1)-1}/S^1$ which generalizes a 2-dimensional Bloch sphere $S^2$ when $d=3$. 
\end{con}
Definitely, the general problem deserves thorough analysis which we postpone for the future work. Our analysis raises also several interesting questions: is the map $\tau_{n,k}$ for $GCD(n,k)=1$ not only optimal but also extremal? Recall, that the Choi map $\tau_{3,1}$ is extremal. However, being extremal it is not exposed \cite{Q12,Q14}. In the recent paper \cite{aniLAA} it was shown that maps $\tau_{n,k}$ with $GCD(n,k)=1$ do not have spanning property which is necessary for exposedness and hence even if they are extremal they can not be exposed.  A similar problem may be posed for the optimized map $\tau_{n,k} - \lambda_{|v\rangle} H_{|v\rangle}$.

%already extremal? Interestingly, the Choi map $\tau_{n,k}$ being extremal is not exposed. It would be interesting to provide the corresponding analysis of exposedness for  $\tau_{n,k}$ with $GCD(n,k)=1$. Actually, if we are able to prove that these maps are exposed then clearly the extremality follows. Recall, that for exposedness  the spanning property  (\ref{SPAN}) is now not sufficient but necessary. 

\section*{Acknowledgements}
The work was supported by the Polish National Science Centre project No. 2018/30/A/ST2/00837.

\appendix

\section{Proof of Theorem \ref{main_th}}
\label{appA}

Let us rewrite the map $\tau_{n,k}(X)$ explicitly in the following matrix form
\begin{eqnarray}
\tau_{n,k}(X) =  {\rm D}(X) - X ,
%&\begin{bmatrix}
%    (n-k) X_{00}+X_{11}+\ldots+X_{kk} & . & . & \dots  & . \\
%    . &  (n-k) X_{11}+X_{22}+\ldots+X_{k+1,k+1} & . & \dots  & . \\
%    \vdots & \vdots & \vdots & \ddots & \vdots \\
%    . & . & . & \dots  &  (n-k) X_{n-1,n-1}+X_{00}+\ldots+X_{k-1,k-1}
%\end{bmatrix}-X. \nonumber\\
\label{map1}
\end{eqnarray}
where ${\rm D}(X) =  {\rm Diag}(D_{0},D_{1},\ldots,D_{n-1})$ is a diagonal matrix such that
\begin{equation}
    D_{i} = (n-k) X_{ii} + X_{i+1,i+1} + \ldots + X_{i+k,i+k} . 
\end{equation}
Let us first observe that for $X = \ket{\psi}\bra{\psi}$, the operator $P_{v_1} \circ X$ is equal to $\ket{\widetilde\psi}\bra{\widetilde\psi}$, where $\widetilde\psi_i = (-1)^i \psi_i$.

Without loss of generality, we consider the action of the map in \eqref{p1}  on the rank 1 projector $X=|\psi \rangle \langle \psi|$:
\begin{equation}
\label{p2}
    \tau_{n,k}^{\lambda} (|\psi \rangle \langle \psi|)=D(|\psi \rangle \langle \psi|)-|\psi \rangle \langle \psi|- \frac{\lambda}{n} |\widetilde{\psi} \rangle \langle \widetilde{\psi}|.
\end{equation}
  Since the above equation consists of the vectors $\{|\psi\rangle,\widetilde{|\psi\rangle}\}$, then it is enough to consider the subspace spanned by these vectors 
  $\{|\psi\rangle,\widetilde{|\psi\rangle}\}$ and show that 
  \begin{equation}
  \label{p3}
      \langle \gamma^* \psi +\delta^* \widetilde{\psi}|\Big(D(|\psi \rangle \langle \psi|)-|\psi \rangle \langle \psi|- \frac{\lambda}{n} |\widetilde{\psi} \rangle \langle \widetilde{\psi}|\Big)|\gamma \psi +\delta \widetilde{\psi} \rangle \geq 0,
  \end{equation}
with $\gamma, \delta \in \mathbb{C}$.
% s.t. $|\gamma|^2+|\delta|^2=1$.
The above equation \eqref{p3} can be explicitly written as
\begin{eqnarray}
\label{p4}
\gamma^* \gamma~\Big(\langle \psi|D|\psi \rangle-|\psi|^4-\frac{\lambda}{n}|\langle\psi|\widetilde \psi \rangle|^2\Big) +
\gamma^* \delta~\Big(\langle \psi|D|\widetilde{\psi} \rangle-|\psi|^2 \langle \psi|\widetilde \psi \rangle -\frac{\lambda}{n} 
\langle \psi|\widetilde \psi\rangle |\widetilde \psi|^2
\Big) + \nonumber\\
\gamma \delta^*~\Big(\langle \widetilde{\psi}|D|\psi \rangle-|\widetilde{\psi}|^2 \langle \widetilde{\psi}| \psi \rangle -\frac{\lambda}{n} 
\langle \widetilde{\psi}| \psi\rangle |\widetilde \psi|^2
\Big) +
\delta^* \delta~\Big(\langle \widetilde \psi|D| \widetilde \psi \rangle-|\langle \psi|\widetilde \psi \rangle|^2-\frac{\lambda}{n} |\widetilde \psi|^4\Big) \geq 0 \nonumber\\
\implies 
\left[
\begin{array}{cc}
 \gamma^*  & \delta^*  \\
\end{array}
\right]  
\left[
\begin{array}{cc}
 \langle \psi|D|\psi \rangle-|\psi|^4-\frac{\lambda}{n}|\langle\psi|\widetilde \psi \rangle|^2  & \langle \psi|D|\widetilde{\psi} \rangle-|\psi|^2 \langle \psi|\widetilde \psi \rangle -\frac{\lambda}{n} 
\langle \psi|\widetilde \psi\rangle |\widetilde \psi|^2  \\
 \langle \widetilde{\psi}|D|\psi \rangle-|\widetilde{\psi}|^2 \langle \widetilde{\psi}| \psi \rangle -\frac{\lambda}{n} 
\langle \widetilde{\psi}| \psi\rangle |\widetilde \psi|^2 & \langle \widetilde \psi|D| \widetilde \psi \rangle-|\langle \psi|\widetilde \psi \rangle|^2-\frac{\lambda}{n} |\widetilde \psi|^4 \\
\end{array}
\right]  
\left[
\begin{array}{c}
 \gamma  \\
 \delta  \\
\end{array}
\right] \geq 0.
\end{eqnarray}
To simplify our calculation, from now on we consider the following notation:
\begin{eqnarray}
     \vec{x}_{od}&=&(|\psi_1|^2,|\psi_3|^2,\ldots, |\psi_{n-1}|^2),~
     \vec{x}_{ev}=(|\psi_0|^2,|\psi_2|^2,\ldots, |\psi_n|^2),~\nonumber\\
 D_{od} &=& \sum_{i=odd} D_i |\psi_i|^2,~
 D_{ev} =\sum_{i=even} D_i |\psi_i|^2, \nonumber\\
  X_{od} &=& \sum_{i=odd} |\psi_i|^2 = \langle \mathbbm{1} | \vec{x}_{od} \rangle,~ X_{ev}=\sum_{i=even} |\psi_i|^2 = \langle \mathbbm{1} | \vec{x}_{ev} \rangle.
\end{eqnarray}
    Under these notations the co-efficients in \eqref{p4} can be written as
    \begin{eqnarray}
         \langle \psi|D|\psi \rangle= \langle\widetilde \psi|D|\widetilde \psi \rangle =D_{ev}+D_{od},~
         |\psi|^2=  |\widetilde \psi|^2= X_{ev}+X_{od},~\nonumber\\
         \langle\psi|\widetilde \psi \rangle = \langle \widetilde \psi| \psi \rangle =X_{ev}-X_{od},~
          \langle \psi|D|\widetilde \psi \rangle =  \langle \widetilde \psi|D| \psi \rangle = D_{ev}-D_{od}.
    \end{eqnarray}
Now we can rewrite the $2 \times 2$ matrix in \eqref{p4} in the below form
\begin{equation}
\left[
\begin{array}{cc}
D_{ev} + D_{od}-(X_{ev}+X_{od})^2-\frac{\lambda}{n} (X_{ev}-X_{od})^2  & 
D_{ev} - D_{od} - (1+\frac{\lambda}{n}) (X_{ev}^2-X_{od}^2)   \\
D_{ev} - D_{od} - (1+\frac{\lambda}{n}) (X_{ev}^2-X_{od}^2)  &
D_{ev} + D_{od}-(X_{ev}-X_{od})^2-\frac{\lambda}{n} (X_{ev}+X_{od})^2
\end{array}
\right].
\end{equation}
The determinant of this matrix is 
\begin{eqnarray}
 \big[ D_{ev} + D_{od}&-& (1+\frac{\lambda}{n})   (X_{ev}^2+X_{od}^2)\big]^2 - \big [ 2 X_{ev} X_{od} (1-\frac{\lambda}{n})\big]^2 - (D_{ev} - D_{od})^2 \nonumber\\
&+& 2 (1+\frac{\lambda}{n}) (D_{ev} - D_{od}) (X_{ev}^2-X_{od}^2)-
 (1+\frac{\lambda}{n})^2 (X_{ev}^2-X_{od}^2)^2 \nonumber\\
 &=& 4 D_{ev} D_{od}-4 (1+\frac{\lambda}{n}) (D_{ev} X_{od}^2+D_{od} X{ev}^2) + 4 (1+\frac{\lambda}{n})^2 X_{ev}^2 X_{od}^2-4 (1-\frac{\lambda}{n})^2 X_{ev}^2 X_{od}^2 \nonumber\\
 &=& 4\Big( \big[D_{ev}-(1+\frac{\lambda}{n}) X_{ev}^2 \big]   \big[D_{od}-(1+\frac{\lambda}{n}) X_{od}^2 \big] - (1-\frac{\lambda}{n})^2 X_{ev}^2 X_{od}^2 \Big)
\end{eqnarray}
Therefore following \eqref{p4}, we need to show that
\begin{equation}
\label{p5}
    \big[D_{ev}-(1+\frac{\lambda}{n}) X_{ev}^2 \big]   \big[D_{od}-(1+\frac{\lambda}{n}) X_{od}^2 \big] - (1-\frac{\lambda}{n})^2 X_{ev}^2 X_{od}^2 ~\geq~ 0.
\end{equation}
For $\lambda=n-k$, we have $1+\frac{\lambda}{n}=\frac{2n-k}{n}$ and $1-\frac{\lambda}{n}=\frac{k}{n}$. 
%Let us consider $P=\sum_{i=1}^{n/2} |e_i \rangle \langle e_{i+1}|$ be the permutation matrix. Then we define the  three  $(\frac{n}{2} \times \frac{n}{2})$ dimensional matrices % containing $\frac{k}{2}$ 1's 
%in the following way:
%\begin{equation}
%    A=(n-k) \mathbb{I}_{n/2}+\sum_{i=1}^{k/2} P_i,~ B=\sum_{i=0}^{k/2-1} P_i
%, ~C=\sum_{i=1}^{k/2} P_i.
%\end{equation}
%Therefore the above equation \eqref{p5} can be simplified as
%\begin{equation}
%    n(D_{ev}-2Y_{ev}^2) (D_{od}-2Y_{od}^2) + k Y_{od}^2 (D_{ev}-2Y_{ev}^2) %+  k Y_{ev}^2 (D_{od}-2Y_{od}^2) \geq 0.
%\end{equation}
%By using the above matrices, the diagonal terms 
$D_{ev}$ and $D_{od}$ can be explicitly written as
\begin{eqnarray}
\label{diagD}
D_{ev}
= \Vec{x}_{ev}^T A_{n,k} \Vec{x}_{ev}+\Vec{x}_{ev}^T B_{n,k} \Vec{x}_{ev}, \\
D_{od}
= \Vec{x}_{od}^T A_{n,k} \Vec{x}_{od}+\Vec{x}_{od}^T C_{n,k} \Vec{x}_{od}.
\end{eqnarray}
Here the three $(\frac{n}{2} \times \frac{n}{2})$ dimensional matrices take the following form:% containing $\frac{k}{2}$ 1's:
\begin{equation}
    A_{n,k} = (n-k) \mathbf{I}_{n/2}+\sum_{i=1}^{k/2} P_{n/2}^i,~ B_{n,k}=\sum_{i=0}^{k/2-1} P_{n/2}^i,~ C_{n,k}=\sum_{i=1}^{k/2} P_{n/2}^i,
\end{equation}
where  $P_n=\sum_{i=1}^{n} |e_i \rangle \langle e_{i+1}|$ be the permutation matrix (unit cyclic shift).
Now $\vec{x}_{ev}$ and $\vec{x}_{od}$ can be written as
\begin{eqnarray}
    \vec{x}_{ev} &=& \frac{2}{n} \vec{\mathbbm{1}} X_{ev}+\vec u=\frac{2}{n} |\mathbbm{1} \rangle \langle \mathbbm{1}| \vec{x}_{ev}+\vec u,~\nonumber\\
    \vec{x}_{od} &=& \frac{2}{n} \vec{\mathbbm{1}} X_{od}+\vec v=\frac{2}{n} |\mathbbm{1} \rangle \langle \mathbbm{1}| \vec{x}_{od}+\vec v,
\end{eqnarray}
 where $\vec u, \vec v$ are variable components of $\vec{x}_{ev}$ and $\vec{x}_{od}$ respectively. By using the above equation, \eqref{diagD} can be written as
 \begin{eqnarray}
    D_{ev} = \frac {2n-k}n X_{ev}^2 + \frac kn X_{ev}X_{od} 
    + \vec u^T A_{n,k} \vec u + \vec u^T B_{n,k} \vec v, \\
    D_{od} = \frac {2n-k}n X_{od}^2 + \frac kn X_{ev}X_{od} 
    + \vec v^T A_{n,k} \vec v + \vec v^T C_{n,k} \vec u.
    %D_{ev}=\frac{4}{n^2} X_{od}^2 (\frac{n}{2} (n-k)+\frac{k}{2} \frac{n}{2})+d_{diff}
 \end{eqnarray}
 and the equation (\ref{p5}) takes the form:
 \begin{equation}
 \label{p6}
    %\big[D_{ev}+\frac kn  X_{ev}X_{od} \big]   \big[D_{od}+\frac kn  X_{ev}X_{od} \big] - (\frac kn)^2 X_{ev}^2 X_{od}^2 = 
    \frac kn  X_{ev}X_{od} (
    \vec u^T A_{n,k} \vec u + \vec u^T B_{n,k} \vec v + 
    \vec v^T A_{n,k} \vec v + \vec v^T C_{n,k} \vec u
    ) + 
    (\vec u^T A_{n,k} \vec u + \vec u^T B_{n,k} \vec v)
    (\vec v^T A_{n,k} \vec v + \vec v^T C_{n,k} \vec u) 
    ~\geq~ 0.
 \end{equation}
 As $\vec{x}_{ev}$ and $\vec{x}_{od}$ have positive entries, their minimal constant components are $\min \vec u$ and $\min \vec v$ respectively, hence $X_{ev} \ge \frac n2 \min \vec u$ and $X_{od} \ge \frac n2 \min \vec v$. We can now estimate the LHS of the above inequality from below:
 \begin{align}
 \label{p6}
    %\big[D_{ev}+\frac kn  X_{ev}X_{od} \big]   \big[D_{od}+\frac kn  X_{ev}X_{od} \big] - (\frac kn)^2 X_{ev}^2 X_{od}^2 = 
    \frac {nk}4  \min \vec u \min \vec v 
    &
    (
    \vec u^T A_{n,k} \vec u + \vec u^T B_{n,k} \vec v
    + 
    \vec v^T A_{n,k} \vec v + \vec v^T C_{n,k} \vec u
    ) \nonumber \\
    & + 
    (\vec u^T A_{n,k} \vec u + \vec u^T B_{n,k} \vec v)
    (\vec v^T A_{n,k} \vec v + \vec v^T C_{n,k} \vec u) 
    ~\geq~ 0.
 \end{align}
 Hence the positivity of the map (\ref{p1}) is equivalent to:
 \begin{align}
 \label{p7}
\forall \vec u, \vec v \in \mathbbm{1}_{n/2}^{\perp},~ \ 
\frac {nk}4  \min \vec u \min \vec v 
    &
    (
    \vec u^T A_{n,k} \vec u + \vec u^T B_{n,k} \vec v
    + 
    \vec v^T A_{n,k} \vec v + \vec v^T C_{n,k} \vec u
    ) \nonumber \\
    & + 
    (\vec u^T A_{n,k} \vec u + \vec u^T B_{n,k} \vec v)
    (\vec v^T A_{n,k} \vec v + \vec v^T C_{n,k} \vec u) 
    ~\geq~ 0.
 \end{align}
\hfill $\Box$

\section{Proof of Proposition \ref{lemma_cj}}
\label{appB}

 Using the lemma \ref{Darek's}, one can estimate the LHS of the inequality (\ref{p7M}):
 \begin{align}
    \label{p8}
    \frac {nk}4  \min \vec u \min \vec v 
    &
    (
    \vec u^T A_{n,k} u+ \vec u^T B_{n,k} \vec v
    + 
    \vec v^T A_{n,k} \vec v + \vec{d}_{od}^T C_{n,k} \vec{d}_{ev}
    ) \nonumber \\
    & + 
    (\vec u^T A_{n,k} \vec u + \vec u^T B_{n,k} \vec v)
    (\vec v^T A_{n,k} \vec v + \vec v^T C_{n,k} \vec u) 
    ~\geq~ \nonumber \\
    \frac {nk}4  \min \vec u \min \vec v \ \cdot \ 
    & \Vec v^T C_{n,k} u + \vec u^T A_{n,k} \vec u \cdot \vec v^T A_{n,k} \vec v
    ~\geq~ 0.
 \end{align}
Let $a_{n,k}^{min}$ be the lowest eigenvalue of $(A_{n,k}+A_{n,k}^T)/2$. We will use the fact, that $x^T A_{n,k} x \ge a_{n,k}^{min} ||x||^2$ and $\forall x \in \mathbbm{1}_{n/2}^\perp \ ||x|| \ge \sqrt{\frac n{n-2}} \min x$, hence $x^T A_{n,k} x \ge a_{n,k}^{min} \frac n{n-2} (\min x)^2$. Using this observation we further estimate the LHS of (\ref{p8}):
\begin{align}
    &
    \frac {nk}4  \min \vec u \min \vec v
    \Vec{d}_{od}^T C_{n,k} \vec u + \vec u^T A_{n,k} \vec u \vec v^T A_{n,k} \vec v \ge
    \nonumber \\
    &
    \frac {nk}4  \min \vec u \min \vec v
    \vec v^T C_{n,k} \vec u + \frac 12 a_{n,k}^{min} \frac n{n-2} (
    (\min \vec v)^2 \vec u^T A_{n,k} \vec u
    +
    (\min \vec u)^2 \vec v^T A_{n,k} \vec v
    )
    =
    \nonumber \\
    &
    \frac 12
    \left[ \begin{array}{cc} (\min \vec v) \vec u^T & (\min \vec u) \vec v^T \end{array} \right]
    \left[ \begin{array}{rr} a_{n,k}^{min} \frac n{n-2} \frac {A_{n,k}+A_{n,k}^T}2 & \frac {nk}4 \frac {C_{n,k}+C_{n,k}^T}2 \\ \frac {nk}4 \frac {C_{n,k}+C_{n,k}^T}2 & a_{n,k}^{min} \frac n{n-2} \frac {A_{n,k}+A_{n,k}^T}2 \end{array} \right]
    \left[ \begin{array}{c} (\min \vec v) \vec u \\ (\min \vec u) \vec v \end{array} \right]
    =
    \nonumber \\
    &
    \frac 12
    \left[ \begin{array}{cc} (\min \vec v) \vec u^T & (\min \vec u) \vec v^T \end{array} \right]
    \left[ \begin{array}{cc} a_{n,k}^{min} \frac n{n-2} \left( (n-k) \mathbf{I}_{n/2} + \frac {C_{n,k}+C_{n,k}^T}2 \right) & \frac {nk}4 \frac {C_{n,k}+C_{n,k}^T}2 \\ \frac {nk}4 \frac {C_{n,k}+C_{n,k}^T}2 & a_{n,k}^{min} \frac n{n-2} \left( (n-k)\mathbf{I}_{n/2} + \frac {C_{n,k}+C_{n,k}^T}2 \right) \end{array} \right]
    \left[ \begin{array}{c} (\min \vec v) \vec u \\ (\min \vec u) \vec v \end{array} \right]
    .
\end{align}
Observe, that $A_{n,k} = \mathbf{I}_{n/2} + C_{n,k}$ and hence $a_{n,k}^{min} = (n-k+c_{n,k}^{min})$. 
%where $c_{n,k}^{(j)}$ is the $j$-th eigenvalue of $(C_{n,k}+C_{n,k}^T)/2$ defined in (\ref{cnkj}).
We will prove, that matrix of the above quadratic form is positive semidefinite. One can simultaneously diagonalise blocks of the matrix making the matrix block-diagonal, with $2 \times 2$ diagonal blocks of the form:
\begin{equation}
    \left[ \begin{array}{cc} 
    \frac n{n-2} (n-k + c_{n,k}^{min})(n-k + c_{n,k}^{(j)}) & \frac {nk}4 c_{n,k}^{(j)} \\ 
    \frac {nk}4 c_{n,k}^{(j)} & \frac n{n-2} (n-k + c_{n,k}^{min})(n-k + c_{n,k}^{(j)}) 
    \end{array} \right].
\end{equation}
We want to prove, that all the blocks are positive, hence:
\begin{equation}
\label{eqc}
    \forall j \in \{1, \dots, \frac{n}{2}-1, \} 
    \ \ 
    (n-k + c_{n,k}^{min})(n-k + c_{n,k}^{(j)}) \ge \frac {(n-2)k}4 |c_{n,k}^{(j)}|
\end{equation}

Now we consider the scenario for $k=2$. Then Eq.~\eqref{eqc} becomes
\begin{equation}
     (n-2 + c_{n,2}^{min})(n-2 + c_{n,2}^{(j)}) \ge \frac {n-2}2 |c_{n,2}^{(j)}|
\end{equation}
One has: $c_{n,2}^{(j)} = \cos(\frac{4\pi}n j) \in [-1,1]$ and hence for $n\ge 4$:
\begin{equation}
    (n-2 + c_{n,2}^{min})(n-2 + c_{n,2}^{(j)}) \ge (n-3)^2 \ge \frac{n-2}2 \ge \frac{n-2}2 |c_{n,2}^{(j)}|
\end{equation}
and the condition (\ref{eqc}) is satisfied.

Now we focus on the case where $k\geq 4$. We have either $c_{n,k}^{(j)}<0$ or $c_{n,k}^{(j)}>0$.

\underline{ \bf{ Case I when $c_{n,k}^{(j)}>0$}}: In this case, Eq.~\eqref{eqc} becomes
\begin{eqnarray}
\label{c11}
     (n-k + c_{n,k}^{min})(n-k + c_{n,k}^{(j)}) \ge \frac {(n-2)k}4 c_{n,k}^{(j)} \nonumber\\
 \implies     \Big((n-k + c_{n,k}^{min})-\frac {(n-2)k}{4}\Big) c_{n,k}^{(j)}+(n-k + c_{n,k}^{min}) (n-k) \geq 0.
\end{eqnarray}
While $c_{n,k}^{min} < 0$, obviously $c_{n,k}^{min} \leq \frac {(n-2)k}{4}-n+k =\frac{(k-4)n+2k}{4}$ for $k\geq 4$. Hence $(n-k + c_{n,k}^{min})-\frac {(n-2)k}{4} < 0$ and the LHS of (\ref{c11}) is a decreasing function of $c_{n,k}^{(j)}$ 
%Therefore the term  $\Big((n-k + c_{n,k}^{min})-\frac {(n-2)k}{4}\Big)$ is negative but $c_{n,k}^{(j)}$ is positive. 
%Hence, the LHS of Eq. \eqref{c11} is an increasing function of $c_{n,k}^{(j)}$ 
 and it is enough to check  \eqref{c11} for $c_{n,k}^{max}$.
 This implies 
 \begin{equation}
     (n-k-c_{n,k}^{min})(n-k-c_{n,k}^{max}) \ge \frac k4 (n-2) c_{n-k}^{max}.
 \end{equation}

  \underline{ \bf{ Case II when $c_{n,k}^{(j)}<0$}}: In this case, Eq.~\eqref{eqc} becomes
\begin{eqnarray}
\label{case2}
     (n-k + c_{n,k}^{min})(n-k + c_{n,k}^{(j)}) \ge -\frac {(n-2)k}4 c_{n,k}^{(j)} \nonumber\\
 \implies     \big(n-k + c_{n,k}^{min}+\frac {(n-2)k}{4}\big) c_{n,k}^{(j)}+ (n-k + c_{n,k}^{min}) (n-k) \geq 0.
\end{eqnarray}
By using the fact~\ref{fact1}, we have
\begin{equation}
    c_{n,k}^{min} \geq -\frac 12 \big(\mu (k+1) + 1\big) \geq -\frac{k+3}{4} \geq -\big(\frac{k^2}{4}+2\big) \geq -n+k-\frac{(n-2)k}{4}, ~\mbox{as}~ n\geq k+2.
\end{equation}
Hence $n-k + c_{n,k}^{min}+\frac {(n-2)k}{4} \geq 0$ and the LHS of \eqref{case2} is a increasing function of  $c_{n,k}^{(j)}$ and it is enough to check \eqref{case2} for $c_{n,k}^{min}$. This implies
\begin{equation}
     (n-k+c_{n,k}^{min})^2 \ge - \frac k4 (n-2) c_{n,k}^{min}.
\end{equation}

%$ \ge \min_{x\in [0,\pi]} \frac 12 (\frac{\sin((k+1)x)}{\sin(x)} - 1) \ge -\frac k2 - 1$, hence $c_{n,k}^{min} \ge -n-\frac k4 (n-6)$ and then:
%To show the first co-efficient in the above  $n-k + c_{n,k}^{min}+\frac {(n-2)k}{4} \geq 0$ as $-c_{n,k}^{min} \leq n \leq n+\frac{k}{4} (n-6)$.
%If $ c_{n,k}^{(j)}=c_{n,k}^{min}$, then Eq.~\eqref{eqc} becomes
%\begin{eqnarray}
 %   (n-k + c_{n,k}^{min})(n-k + c_{n,k}^{(j)}) + \frac {(n-2)k}4 c_{n,k}^{(j)} \ge 
 %   (n-k + c_{n,k}^{min})^2+ \frac {(n-2)k}4 c_{n,k}^{(min)} \geq 0 \nonumber\\
 %   \implies   (n-k + c_{n,k}^{min})^2+  (n-k + c_{n,k}^{min}) \frac{(n-2)k}{4}-(n-k) \frac{(n-2)k}{4} \geq 0 \nonumber\\
   % \implies  n-k + c_{n,k}^{min} \leq \frac{1}{2} \Big(-\frac{(n-2)k}{4} \pm \sqrt{\frac{(n-2)^2 k^2}{16}+(n-k) (n-2)k} \Big) \nonumber\\
  % \implies -(n-k)-\frac{(n-2)k}{8} \Big(1+\sqrt{1+\frac{16 (n-k)}{(n-2) k}} \Big) \leq  c_{n,k}^{min} \leq -(n-k)-\frac{(n-2)k}{8} \Big(1-\sqrt{1+\frac{16 (n-k)}{(n-2) k}} \Big) \nonumber\\
%\end{eqnarray}

% If $k=2$, then the  LHS of Eq. \eqref{c11} is non-increasing function of  $ c_{n,k}^{(j)}$, and it is enough to check for the smallest positive  $ c_{n,k}^{(j)}$. Clearly for $k=2$, LHS of Eq. \eqref{c11} is negative, but the RHS of  Eq. \eqref{c11} is positive, hence  Eq. \eqref{c11} is always satisfied for $k=2~\forall n$.

\hfill $\Box$

\section{Proof of Corollary \ref{COR1}}
\label{AppC}
\begin{itemize}
    \item $n=2k$: In this case, we find $c_{n,k}^{max}=0$ and $c_{n,k}^{min}=-1$. Then from \eqref{c_max_cond} and \eqref{c_min_cond}, one has
    \begin{equation}
         (n-k+c_{n,k}^{min})(n-k+c_{n,k}^{max}) - \frac k4 (n-2) c_{n,k}^{max}=(2k-k-1) (2k-k)=k (k-1) >0,
    \end{equation}
    and
    \begin{equation}
        (n-k+c_{n,k}^{min})^2  + \frac k4 (n-2) c_{n,k}^{min}=(2k-k-1)^2-\frac k4 (2k-2)=\frac{1}{2} (k-1) (k-2) \geq 0.
    \end{equation}

    \item $n=2k+2$: Here all $c_{n,k}$'s are same, and therefore $c_{n,k}^{min}=-\frac{1}{2}$.
Clearly,  from  \eqref{c_min_cond}, 
    \begin{equation}
        (n-k+c_{n,k}^{min})^2  + \frac k4 (n-2) c_{n,k}^{min}=(2k+2-k-\frac 12)^2 -\frac{k}{8} (2k+2-2)=(k+\frac 32)^2-\frac{k^2}{4}=\frac 34 (k+1) (k+3) \geq 0.
    \end{equation}

     \item $n=2k+4$: In this scenario, we find $c_{n,k}^{max}=0$ and $c_{n,k}^{min}=-1$. Then from \eqref{c_max_cond} and \eqref{c_min_cond}, one has
    \begin{equation}
         (n-k+c_{n,k}^{min})(n-k+c_{n,k}^{max}) - \frac k4 (n-2) c_{n,k}^{max}=(2k+4-k-1) (2k+4-k)=
         (k+3) (k+4) >0,
    \end{equation}
    and
    \begin{equation}
        (n-k+c_{n,k}^{min})^2  + \frac k4 (n-2) c_{n,k}^{min}=(2k+4-k-1)^2-\frac k4 (2k+4-2)=(k+3)^2-\frac{k}{2} (k+1)=\frac 12 (k^2+11k+9) \geq 0.
    \end{equation}

\end{itemize}

\section{Proof of Corollary \ref{COR2}}
\label{AppD}
 For $k \ge 4$ one has:
\begin{equation} \label{cor_ineq}
    16 n \ge 2k^2+8(\mu+4)k+4(\mu+1)-27    
    \ge (2\mu k^2 + 2(9\mu+17) k +8\mu - 41
    \ge  8 ((2 + \mu)k + \mu + 1). 
\end{equation}
Hence under the assumption, one has that $\frac 12 (\mu (k+1) + 1) \le n-k$. Then $-c_{n,k}^{min} \le n-k$ and $(n-k+c_{n,k}^2)^2 \ge (n-k-\frac 12 (\mu(k+1)+1))^2$, and:
\begin{align}
    (n-k+c_{n,k}^{min})^2 
    & + \frac k4(n-2)c_{n,k}^{min} 
    \ge
    (n-k-\frac 12 (\mu k + (\mu+1)) )^2 - \frac 18 (\mu k^2+(\mu+1)k)(n-2)
    \nonumber \\
    & = n^2 - \frac 18 (\mu k^2+(9\mu+17)k+8(\mu+1))n
    + \frac 14 ((\mu^2+5\mu+4)k^2+(2\mu^2+7\mu+5)k+(\mu+1)^2)
    \nonumber \\
    & = (n - \frac 1{16} (\mu k^2+(9\mu+17)k+8(\mu+1)))^2 
    \nonumber \\
    & - \frac 1{16^2}
    (\mu^2 k^4 + 2(9\mu+17)\mu k^3 + (33\mu^2+2\mu+33)k^2+16(\mu^2-2\mu-3)k)
    \nonumber \\
    & \ge (n - \frac 1{16} (\mu k^2+(9\mu+17)k+8(\mu+1)))^2 
    \nonumber \\
    & - \frac 1{16^2}
    (\mu^2 k^4 + 2(9\mu+17)\mu k^3 + (33\mu^2+2\mu+33)k^2+16(\mu^2-2\mu-3)k
    \nonumber \\
    & + 2(24\mu^2-103\mu+128) (k-4)^2 
    + 2(1+\mu)(215+184\mu)(k-4) 
    + (704\mu^2-104\mu+25))
    \nonumber \\
    & = (n - \frac 1{32} (k^2+43k-24))^2 - \frac 1{16^2} (\mu k^2+(9\mu+17)k-49)^2
    \nonumber \\
    & = \frac 1{16^2} (16n - 8\mu - 57)(16n - 2\mu k^2-2(9\mu+17)k+8\mu-41), 
    %\nonumber \\
    %& \ge \frac 1{16^2} (16n - 8\mu - 57)(16n - .7k^2-41k+38)
    %\ge 0
    %\nonumber \\
    %& \Leftrightarrow \ n \ge (2k^2+86k-142)/32
\end{align}
and the RHS is non-negative due to the assumption and (\ref{cor_ineq}) and the condition holds. Similarly:
\begin{align}
    (n-k+c_{n,k}^{min})(n-k+c_{n,k}^{max}) 
    & - \frac k4(n-2)c_{n,k}^{max}
    \ge (n-k-\frac 12 (\mu(k+1)+1)) (n-k) - \frac{k^2}8 (n-2) 
    \nonumber \\
    & = n^2-n \frac 18 (k^2+4(\mu+4)k+4(\mu+1)) + (\frac {5+2\mu}4k^2+\frac{\mu+1}2 k)
    \nonumber \\
    & = (n - \frac 1{16} (k^2+4(\mu+4)k+4(\mu+1)))^2 
    \nonumber \\
    & - \frac 1{16^2} (k^4 + 8(\mu+4)k^3 + 8(2\mu^2+\mu-7)k^2 + 32\mu(\mu+1)k 
    + 16(\mu+1)^2)
    \nonumber \\
    & \ge (n - \frac 1{16} (k^2+4(\mu+4)k+4(\mu+1)))^2 
    - \frac 1{16^2} \Big(k^4 + 8(\mu+4)k^3 + 8(2\mu^2+\mu-7)k^2 
    \nonumber \\
    & + 32\mu(\mu+1)k + 16(\mu+1)^2
    \nonumber \\
    & + 2(119+60\mu) (k-4)^2 + 8(-4\mu^2+79\mu+90)(k-4) 
    \nonumber \\
    & + (-55+384\mu-144\mu^2)
    \Big)
    \nonumber \\
    & = (n - \frac 1{16} (k^2+4(\mu+4)k+4(\mu+1)))^2 
    - \frac 1{16^2} (k^2+4(4+\mu)k-37)^2
    \nonumber \\
    & = \frac 1{256} (16n - 2k^2 - 8(\mu+4)k - 4(\mu+1) + 27)(16n - 4(\mu+1)-27),
    %\ge 0
    %\nonumber \\
    %& \Leftrightarrow \ n \ge (2k^2+36k-33)/16
\end{align}
and again it is non-negative due to the assumption and (\ref{cor_ineq}) and the condition holds. $\square$

\section{Nummerical calculations}
\label{AppE}

The code is placed in the repository: \texttt{https://github.com/gniewko-s/Optimisation\_GCD\_3}. The code \texttt{opt.py} produces the data and serialise it to files: \texttt{data\_6\_3.pi}, \texttt{data\_9\_3.pi} and \texttt{data\_9\_6.pi}. The code \texttt{plot\_data.py} uses the serialised data to create the subfigures of the Figure \ref{fig1}.

The code \texttt{opt.py} defines the functions as follows: The function \texttt{S(n,k)} returns the circulant matrix $(n-k)I + \sum_{i=1}^{k/2} S^i$ for given $n,k$ (where $S$ is a unit-shift matrix defined in (\ref{unit_shift})). 
It is used in the subsequent function \texttt{mapp(n,k,phi,theta)} returning a map: $\lambda \mapsto (\mathbb{R}^n \ni X \mapsto \det \tau_{n,k}^{\lambda,\alpha,\beta}(XX^\dagger))$, where $\alpha = \cos(\theta/2)$ and $\beta = \cos(\theta/2) \exp(i\phi)$. 
A simple observation that changing phases of coordinates of $X$ results in a unitarily equivalent result let us consider only real $X$ and reduce the number of optimisation parameters. 
This function is minimised over $X$ in the function \texttt{robust\_min}, returning minimum over \texttt{M} successful minimisations by use of \texttt{scipy.optimise.minimize} function. 
We are interested in finding the maximal value of $\lambda$ for which the minimisation gives $0$ (what corresponds to a singular matrix). 
We obtain it by a Newton method, starting from initial points $\lambda = n$ and $\lambda = n-1$. 
It is implemented in the function \texttt{lmax\_new}. The function \texttt{lmax} wraps it by printing some run-time information. In this way, we are able to calculate $\lambda^{max}_{\ket{v}}$ for any $\ket{v}$ related to a point on the Bloch sphere of spherical coordinates $(\phi,\theta)$.
    
The function \texttt{prepare\_grid} prepares the grids in spherical coordinates ($\phi,\theta$) on the Bloch Ball and the grid of optimisation values. The grid of values is initialised by the maximisation results on the poles and with $n-k$ between the poles. The tuple of three 2-dimensional arrays is then serialised to a file. Next, for each point on the grid, we invoke the function \texttt{calculate\_point}, which loads the data from the file, calculates the $\lambda^{max}_{\ket{v}}$ for a given point, replaces the value in the array of results and serialise the data back to the file.

\end{document}